# Rigorous negative-ion binding energies in low-energy electron elastic collisions with heavy multi-electron atoms and fullerene molecules: validation of electron affinities

Alfred Z. Msezane and Zineb Felfli

Department of Physics and Center for Theoretical Studies of Physical Systems, Clark Atlanta University, Atlanta, Georgia 30314, USA

**Abstract**.
Dramatically sharp resonances manifesting stable negative-ion formation characterize Regge pole-calculated low-energy electron elastic total cross sections (TCSs) of heavy multi-electron systems. The novelty of the Regge pole analysis is in the extraction of rigorous and unambiguous negative-ion binding energies (BEs), corresponding to the measured electron affinities (EAs) of the investigated multi-electron systems. The measured EAs have engendered the crucial question: is the EA of multi-electron atoms and fullerene molecules identified with the BE of the attached electron in the ground, metastable or excited state of the formed negative-ion during collision? Inconsistencies in the meaning of the measured EAs are elucidated.

## 1. Introduction

In the electron impact energy range $0.0 \leq E \leq 10.0$ eV, dramatically sharp resonances manifesting stable ground, metastable and excited negative-ion formation, shape resonances (SRs) and Ramsauer–Townsend (R-T) minima characterize the Regge pole-calculated low-energy electron elastic total cross sections (TCSs) of heavy multi-electron atoms and fullerene molecules [1]. The energy positions of the sharp resonances correspond to the measured electron affinities (EAs) of the considered multi-electron atoms and fullerene molecules. Indeed, the extraction from the TCSs of rigorous and unambiguous negative-ion binding energies (BEs), the SRs and the R-T minima, without any experimental or other theoretical assistance demonstrates the novelty and strength of the Regge pole analysis and its vital importance in the understanding of low-energy electron collisions with complex multi-electron systems through negative-ion formation.

The recent theoretical investigation of low-energy electron elastic collisions with heavy multi-electron atoms and fullerene molecules, using Regge pole analysis, also discussed the meaning of the measured EA within two prevailing contexts [1]. The first viewpoint considers the EA to correspond to the electron binding energy when it is attached in the ground state of the formed negative-ion during the collision. The second view interprets the EA as corresponding to the BE of the attached electron in an excited state of the formed negative-ion. Examples of the first case are the measured EAs of the Au, Pt and the highly radioactive At atoms [2-7] as well as of the $C_{60}$ and $C_{70}$ fullerene molecules [8-12]. The measured EAs of Nb [13, 14], Hf [15], the lanthanide atoms Eu [16, 17] and Tm [18] and the actinide atoms Th [19] and U [20, 21] as well as the theoretical EAs of Am, Bk, Cf, Fm and Lr [22-26] represent the second interpretation of the EA. Clearly, the unambiguous and confusing interpretation of the experimental and/or theoretical EAs of the heavy multi-electron atoms and fullerene molecules still continues to plague the meaning of the EAs. Indeed, the lack of uniformity in the meaning of the measured EAs of these multi-electron systems, now proliferating in the literature has engendered confusion and motivated this investigation. It must be stressed, however that both viewpoints do find interpretation in the Regge pole-calculated rigorous negative-ion BEs.

To clarity our objective further, we demonstrate the second viewpoint of the meaning of the EA by using the Nb atom as an example. The EA of atomic Nb is both interesting and revealing because there are two measured EA values, namely 0.917eV[13] and 0.894eV[14] as well as two theoretical EAs 0.82eV[27] and 0.99eV[28]. Their interpretation notwithstanding, the agreement among these values is quite good and with the Regge pole-calculated metastable BE value of 0.902eV. Notably, the calculation of the EA of Nb in [27] used a gradient-corrected exchange correlation functional and a Nb scalar-relativistic core [27]. These results for Nb further demonstrate the great need to ascertain precisely the state whence the photodetachment process originates.

Experimental investigation of the EAs of the lanthanide atoms is challenging because of the difficulty of producing sufficient negative ions for use in photodetachment experiments [16]. And for these atoms problems concerning the interpretation of what is meant by the measured EAs have already been discussed in [29, 30]; to our knowledge, they have not yet been resolved. Further difficulties are encountered with the lanthanide atoms because the structures of their TCSs are significantly different from those of the standard atoms and fullerenes considered in this paper; compare for example Figs. 1 and 3 here. Experimental investigations of the actinide atoms are limited by the difficulty of handling them because they are highly radioactive. In [1] the sensitivity of the Regge pole-calculated R-T minima and SRs to the electronic structure and dynamics of the Bk and Cf actinide atoms was demonstrated and utilized as novel and rigorous validation of the experimental observation, using a nanogram of material [31]. The experiment identified a weak spin-orbit-coupling in atomic Bk while a jj coupling scheme described atomic Cf, thereby strengthening the conclusion that Cf is a transitional element in the actinide series [31]. Significantly, this experimental breakthrough [31] and the recent first ever EA measurements of the highly radioactive elements At [7], Th [19] and U[20, 21] represent significant advances in the measurements of the challenging to handle highly radioactive elements. And more such measurements in other radioactive atoms can be expected in the near future. Of great concern and puzzle, however is that the measured [7] and the calculated [32-35] and references therein EAs of At correspond to the ground state BE of the formed At⁻ anion during the collision, while the measured EAs of Th and U are identified with the BEs of the attached electron in the excited states of the formed negative-ions during collision. Consequently, reliable theoretical predictions and guidance are essential for a fundamental understanding and interpretation of what is actually being measured.

Theoretically, most of the methods developed in atomic physics, were designed with the fundamental primordial task of reproducing experimental results with high accuracy, but not necessarily for reliable prediction. This effort yielded very sophisticated methods for the most part, that provided agreement with experiment, but did very little to unravel and elucidate the fundamental physics at play, particularly at low-electron energies. The conventional quantum mechanical approach to scattering problems describes the solution as a partial-wave (PW) series where the summation is over the orbital (or total) angular momentum quantum number. The PW expansion, which may contain hundreds, if not thousands of terms, is notoriously slowly convergent. However, if the angular momentum is allowed to become complex-valued, the slowly convergent PW series becomes rapidly converging. This leads us to the concept of Regge poles. Simply put, Regge poles are generalized bound-states, *i.e.* solutions of the Schrödinger equation where the impact energy, E is real, positive and the angular momentum, λ is complex. Mathematically, being the poles of the scattering matrix, S they rigorously define resonances.

The main objective of this paper is to demonstrate and elucidate the problem of the ambiguous meaning of the measured EA now permeating the literature, and its dire need for a rigorous solution, provided here through the anionic BEs extracted from our Regge pole-calculated electron elastic TCSs in the energy range $0.0 \leq E \leq 10.0$ eV. Results for Nb, the large multi-electron atoms (Hf, Au, Pt and At), the lanthanide atoms (Eu and Tm), the actinide atoms (Th, U, Am, Bk, Cf, Fm and Lr), and the $C_{60}$ and $C_{70}$ fullerene molecules are presented. From the TCSs the unambiguous and reliable ground, metastable and excited states negative-ion BEs of the formed anions during the collisions, are extracted and compared with the measured EAs of the investigated atoms and fullerene molecules to understand the current and guide future measurements of the EAs in terms of the rigorous ground, metastable and excited states BEs of the formed negative-ions during the collision.

## 2. Method of Calculation

Here the rigorous Regge pole method has been used to calculate the electron elastic TCSs and hence the anionic BEs. Regge poles, singularities of the S-matrix, rigorously define resonances [36, 37] and in the physical sheets of the complex plane they correspond to bound states [38]. The Regge poles formed during low-energy electron elastic scattering have been confirmed to become stable bound states [39]. In the Regge pole, also known as the complex angular momentum (CAM), method the important and revealing energy-dependent Regge Trajectories are also calculated. Their effective use in low-energy electron scattering has been demonstrated in for example [40, 41]. The near-threshold electron–atom/fullerene collision TCS resulting in negative-ion formation as resonances is calculated using the Mulholland formula [42]. The TCS below fully embeds the essential electron-electron correlation effects [43, 44] (atomic units are used throughout):

$$\sigma_{tot}(E) = 4\pi k^{-2} \int_0^\infty \text{Re}[1 - S(\lambda)]\lambda d\lambda$$
$$- 8\pi^2 k^{-2} \sum_n \text{Im} \frac{\lambda_n \rho_n}{1+\exp(-2\pi i \lambda_n)} + I(E) \quad (1)$$

In Eq. (1) S(λ) is the S-matrix, $k = \sqrt{2mE}$, $m$ being the mass and $E$ the impact energy, $\rho_n$ is the residue of the S-matrix at the $n^{th}$ pole, $\lambda_n$ and $I(E)$ contains the contributions from the integrals along the imaginary λ-axis (λ is the complex angular momentum); its contribution has been demonstrated to be negligible [40].

As in [45] here we consider the incident electron to interact with the complex heavy system without consideration of the complicated details of the electronic structure of the system itself. Therefore, within the Thomas-Fermi theory, Felfli et al [46] generated the robust Avdonina-Belov-Felfli (ABF) potential which embeds the vital core-polarization interaction

$$U(r) = -\frac{Z}{r(1+\alpha Z^{1/3} r)(1+\beta Z^{2/3} r^2)} \quad (2)$$

In Eq. (2) Z is the nuclear charge, α and β are variation parameters. For small $r$, the potential describes Coulomb attraction between an electron and a nucleus, $U(r) \sim -Z/r$, while at large distances it has the appropriate asymptotic behavior, viz. $\sim -1/(\alpha\beta r^4)$ and accounts properly for the polarization interaction at low energies. Notably, for an electron, the source of the bound states giving rise to Regge Trajectories is the attractive Coulomb well it experiences near the nucleus. The addition of the centrifugal term to the well 'squeezes' these states into the continuum [47].

For larger CAM, λ the effective potential develops a barrier. Consequently, a bound state crossing the threshold energy E = 0 in this region may become an excited state or a long-lived metastable state. As a result, the highest "bound state" formed during the collision is identified with the highest excited state, here labeled as EXT-1, see Table 1 for example. As E increases from zero, the second excited state may form with the anionic BE labeled, EXT-2. For the metastable states, similar labeling is used as MS-1, MS-2, etc. However, it should be noted here that the metastable states are labeled relative to the anionic ground state. The CAM methods have the advantage in that the calculations are based on a rigorous definition of resonances, viz. as singularities of the S-matrix [47, 48].

The strength of this extensively studied potential [49, 50] lies in that it has five turning points and four poles connected by four cuts in the complex plane. The presence of the powers of Z as coefficients of $r$ and $r^2$ in Eq. (2) ensures that spherical and non-spherical atoms and fullerenes are correctly treated. Also appropriately treated are small and large systems. The effective potential $V(r) = U(r) + \lambda(\lambda+1)/2r^2$ is considered here as a continuous function of the variables $r$ and complex $\lambda$. The details of the numerical evaluations of the TCSs have been described in [44] and references therein; see also [51].

The calculations of the TCSs, limited to the near-threshold energy region, namely below any excitation threshold to avoid their effects, are obtained by solving the Schrödinger equation as described in [44]. The parameters "$\alpha$" and "$\beta$" of the potential in Equation (2) are varied, and with the optimal value of $\alpha = 0.2$ the $\beta$-parameter is further varied carefully until the dramatically sharp resonance appears in the TCS. This is indicative of stable negative-ion formation during the collision and the energy position matches with the measured EA of the atom/fullerene molecule; see for example the Au and $C_{60}$ fullerene TCSs in [1]. This has been found to be the case in all the atoms and fullerenes we have investigated thus far.

## 3. Results

To better understand and appreciate the problem we are discussing here, we first consider the TCSs of the standard atomic Au and $C_{60}$ fullerene molecule, given in Fig. 1 of Ref. [1]. Clearly seen from the figure is that in the Au TCSs (left panel of Fig. 1) there are three dramatically sharp resonances representing stable negative-ion formation in the ground (2.26eV), metastable (0.832eV) and excited (0.326eV) states of the formed negative ions during the collision. In the $C_{60}$ fullerene TCSs (right panel of Fig. 1) there are five dramatically sharp resonances, corresponding to the ground state BE (2.66eV), two metastable BEs (1.86eV and 1.23eV) and two excited states BEs (0.378eV and 0.203eV). In both cases the measured EAs of the Au atom and the $C_{60}$ fullerene molecule correspond to the anionic BEs of the attached electron in the ground states of the formed anions during the collision and not to those of the metastable or the excited states, see also Table 1. Importantly, delineation of the dramatically sharp resonances in the TCSs of both Au and $C_{60}$ ensures the correct interpretation of what is being measured. It is noted here that both the Au and $C_{60}$ TCSs also abound in SRs and R-T minima. Similarly, for Pt, At and $C_{70}$ the measured EAs correspond to the BEs of the electron when it is attached in the ground states of the formed negative ions (see Table 1 here). Indeed, the excellent agreement between the measured EAs and the Regge pole-calculated ground state anionic BEs of these multi-electron systems gives great credence to the power of the Regge pole analysis to produce rigorous and unambiguous BEs without any assistance from either experiment or any other theory, as well as to our interpretation of the EAs of these complex systems, viz. as corresponding to the ground state BEs of the formed negative ions during the collisions.

Recall that the primary objective of this paper is to subject the measured and/or calculated EAs of the investigated atoms and fullerene molecules in this paper to the Regge pole-calculated ground, metastable and excited states negative-ion BEs of the formed anions during the collisions for unambiguous interpretation of the EAs. For the Au, Pt and At atoms as well as the fullerene molecules $C_{60}$ and $C_{70}$ the measured EAs have been identified with the Regge pole-calculated ground states BEs of the formed negative ions during the collisions. Similarly, the remaining Nb, Hf, the lanthanide (Eu and Tm) and the actinide (Th, U, Am, Bk, Cf, Fm and Lr) atoms will be subjected to a similar process in order to determine the meaning of the measured/calculated EAs in terms of their Regge pole-calculated negative-ion BEs. The Figs. 1-5 of the TCSs demonstrate the clear delineation of the dramatically sharp resonances leading to the rigorous and unambiguous determination of the ground, metastable and excited states BEs of the formed negative ions during the collisions. Table 1 summarizes the BEs of the various atoms and fullerenes, extracted from the TCSs of interest here and compares them with the measured/calculated EAs.

To simplify our discussions of the various atoms we have grouped them for convenience as follows: 3.1 Nb, Eu and Tm atoms because for these atoms the measured EAs are identified mainly with the Regge pole-calculated metastable/excited states BEs. 3.2 Hf atom; the TCSs for Hf resemble those of the lanthanide atoms. 3.3 Th and U atoms; these actinide atoms have both measured and calculated EAs. 3.4 Bk and Cf atoms; the interest in them is that a recent experiment probed their structure and dynamics using a nanogram matter. 3.5 Fm and Lr atoms; being at the end of the actinide series, they have several calculated EAs available to compare with. 3.6 Relativistic effects in electron affinity calculations.

### 3.1  Nb, Eu, Tm atoms

As seen from Table 1, for atomic Nb the measured and the calculated EAs agree very well, and with the Regge pole metastable BE value of 0.902eV and not with the ground or the excited states BEs of 2.48eV and 0.356eV, respectively. Figure 1 displays the typical TCSs of the relatively small, Eu and the large, Tm lanthanide atoms; their behaviors are significantly different from those of Au, the actinide atoms and the fullerene molecules. This explains the difficulty encountered by both experiments and theory in interpreting the measured and the calculated EAs of the lanthanide atoms, thereby calling for careful analysis. For the Eu atom, the latest measured EA (0.116eV) [16] is in outstanding agreement with the Regge pole BE of 0.116eV and with the MCDF-RCI calculated EA value of 0.117eV [22]. But the Regge pole value corresponds to an excited state BE of Eu. Notably, the previously measured EA value (1.053eV) [17] of Eu also agrees excellently with the Regge pole-metastable BE of 1.08eV. Furthermore, the 1.029eV measured EA of atomic Tm [18] and the Regge pole-calculated metastable BE value of 1.02eV also agree excellently, see Fig. 1. Consequently, from the measured EAs of the three atoms Nb, Eu and Tm, it is clear that rigorous BEs are essential for guiding the interpretation of the measured EAs of these atoms. With this, the simple questions follow: 1) Does the measured EA of the Eu atom correspond to the BE of the attached electron in an excited state of the formed anion during the collision? 2) Conversely, are the measured EAs of Eu and Tm atoms identified with the Regge pole-metastable BE values of 1.08eV and 1.02eV, respectively as shown in Fig. 1?

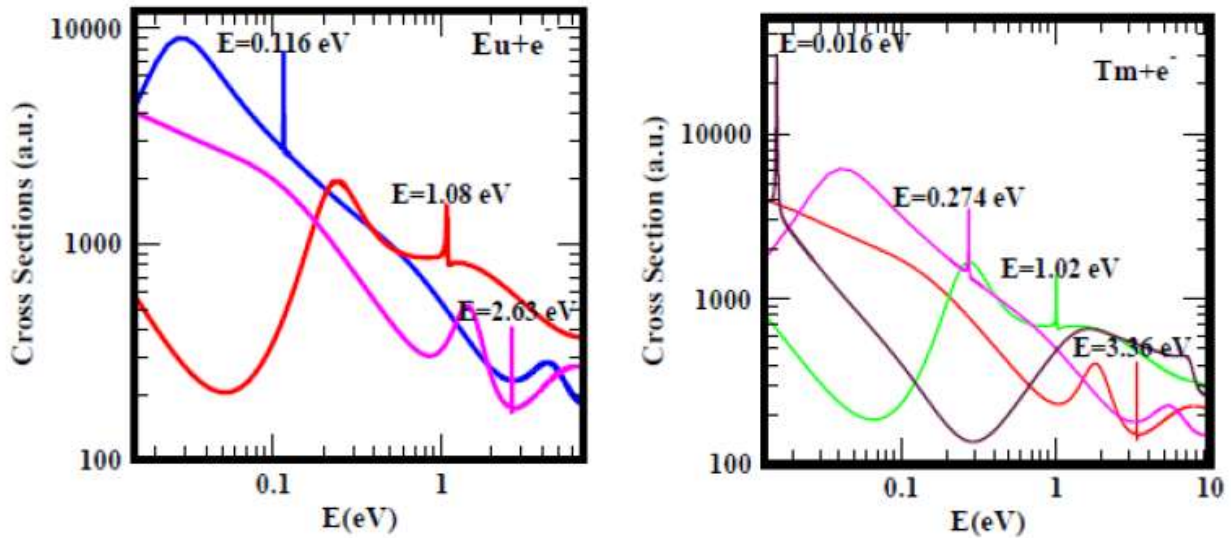

**Figure 1:** Total cross sections (a.u.) for electron elastic scattering from atomic Eu (left panel) and Tm (right panel). For Eu the pink, red and blue curves represent the TCSs for the ground, metastable and the excited states, respectively. For Tm atom the red, green, pink and brown curves represent the TCSs for the ground, metastable and the two excited states, respectively. The dramatically sharp resonances in both figures correspond to the Eu⁻ and Tm⁻ negative-ions formed during the collisions.

## 3.2 Hf atom

Figure 2 depicts the Regge pole-calculated TCSs of atomic Hf; they are characterized by four dramatically sharp peaks, manifesting negative-ion formation in the ground state (BE of 1.68eV), the metastable state (BE of 0.525eV) and the two excited states (BEs of 0.017eV and 0.113eV). There is also a SR at 0.232eV, just before the 0.525eV peak. The MCDF-RCI calculated EA of Hf is 0.114eV [52] and agrees excellently with the Regge pole-calculated excited state BE of 0.113eV [53]. The recently measured EA of Hf is 0.178eV [15] and is relatively close to the MCDF-RCI EA [52] as well as to the Regge pole excited state BE (0.113eV) [53]. Since the measured EA of Hf is close to the RCI EA and the Regge pole BE of an excited state, the measured EA of Hf is identified with the BE of an excited state of Hf contrary to the cases of the Au, Pt and At atoms as well as of the $C_{60}$ and $C_{70}$ fullerene molecules.

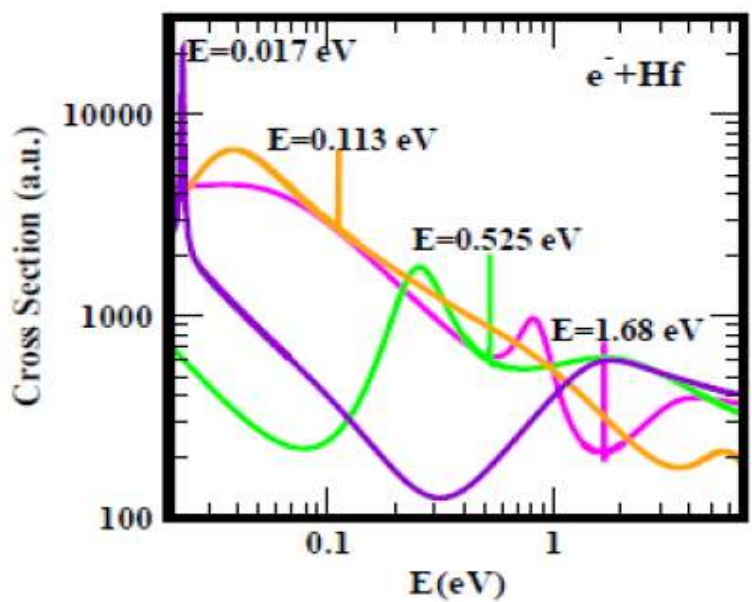

**Figure 2:** Total cross sections (a.u.) for electron elastic scattering from Hf. The pink, green, orange and purple curves represent the TCSs for the ground, metastable and the two excited states, respectively. The dramatically sharp resonances correspond to the Hf⁻ anionic formation during the collisions.

## 3.3 Th and U atoms

For a better understanding of our discussion of the EAs of the actinide atoms, it is important to first place the presentation in perspective. From Table 1, beginning with the results for Au through $C_{70}$, the Regge pole ground states BEs match excellently the measured EAs. Currently, there are only two actinide atoms with measured EAs, viz. Th [19] and U [20, 21]; the latter measurements were published after our guiding paper [54]. However, theoretical EAs are available for the actinide atoms from Th through Lr. Indeed, here we need a standard of reference to sort out the various theoretical EAs which are riddled with uncertainty and lack definitiveness. To understand the actinide data we need to look carefully at the EA values of the highly radioactive At atom. The measured EA [7] of At was supported by various sophisticated theoretical calculations [32-35], including the Regge pole-calculated ground state BE. In [34] an extensive comparison of various theoretical EAs of At has been carried out as well; the EAs vary extensively demonstrating the need for a reliable theory to guide the measurements of the EA.

For Th the measured EA is 0.608eV [19] and the calculated value by the same experiment is 0.599eV [19]. These values are close to the Regge pole SR at 0.61eV and the second excited state BE of 0.549eV. It must be noted here that the highest excited state BE of the Th anion is 0.149eV, not shown in the Figure 3, but is included in Table 1 as well as the anionic ground state BE of the Th anion, 3.09eV.

This is important since the experimenters claimed that their EA corresponded to electron attachment BE in the highest excited state. Also notable here are the EAs calculated using the MCDF-RCI [23] and GW [24], being 0.368eV and 1.17eV, respectively. From Table 1, we can safely conclude that the measured EA of Th [19] is identified with the Regge pole BE of an excited state of the formed negative ion of Th during the collision. Consequently, we do not understand the radical departure from the identification of the EA with the ground state BE of the formed negative ion during the collision as was done for Au, At, Pt and the fullerenes.

Next we consider the measured EAs of U, being 0.315eV [20] and 0.309eV [21], while the calculated EA is 0.232eV [21], also presented in Table 1. These values are close to each other and to the Regge pole BE of the first excited state of the formed $U^-$ anion of 0.220eV, the MCDF-RCI EA value of 0.175eV [55] and to the 0.373eV GW EA [24]. However, the Regge pole BE value of the second excited state, 0.507eV agrees very well with the EA calculated by the GW method 0.53eV [24]. Here we are confronted with the inconsistency in the identification of what is actually being measured. In Th the measured EA is close to the Regge pole-calculated anionic BE of the second excited state, while in U both the measured EAs [20, 21] are close to the BE of the first excited state. It is this inconsistency in the interpretation of the measured EAs with respect to the Regge pole calculated anionic BEs that is puzzling and of concern to us as well. It must be noted here that some of the sophisticated theoretical methods used to guide the measurements are riddled with uncertainties and lack definitiveness. As pointed out previously, these theories tend to calculate the BEs of excited states and identify them with the EAs.

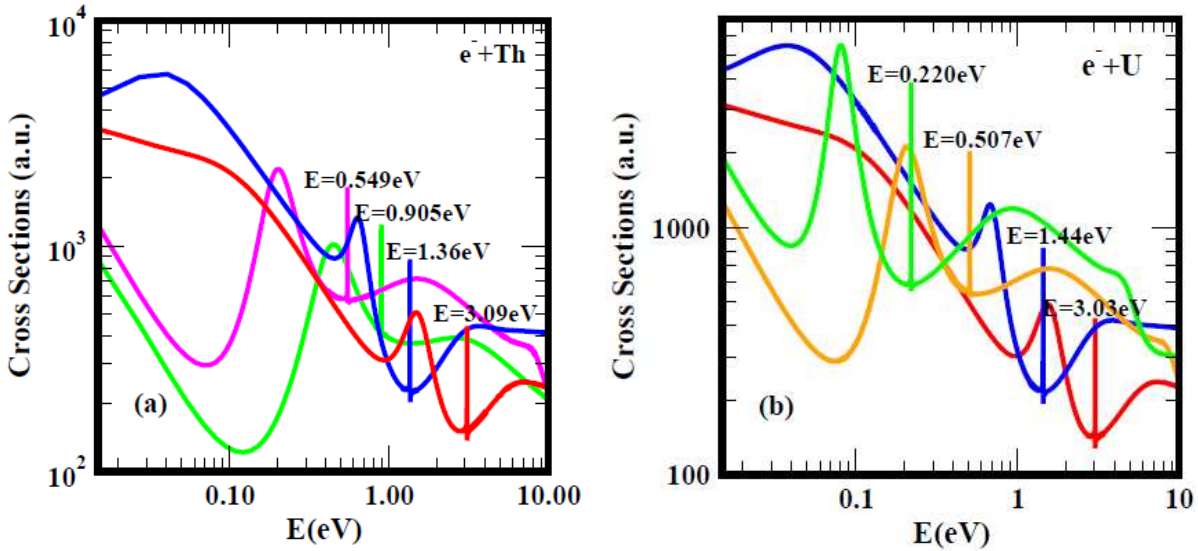

**Figure 3:** Total cross sections (a.u.) for electron elastic scattering from atomic Th (left panel) and U (right panel) are contrasted. For atomic Th the red, blue and green curves represent TCSs for the ground and the two metastable states, respectively while the pink curve corresponds to an excited state TCS. For U the red and the blue curves represent TCSs for the ground and the metastable states, respectively. The orange and the green curves are TCSs for the excited states. The dramatically sharp resonances in both figures correspond to the $Th^-$ and $U^-$ negative-ions formation during the collisions.

### 3.4  Bk and Cf atoms

The proliferation in the literature of ambiguous and confusing EAs of the actinide atoms from Th through Lr has necessitated the careful evaluation/assessment of the measured and/or calculated EAs in the context of the rigorous Regge pole-calculated BEs in order to determine their meaning. As seen from the Table 1, it is particularly difficult to make sense of the existing theoretical EAs of the Am, Bk and Cf actinide atoms.

The selection of the Bk and Cf TCSs shown in Fig. 4 for presentation is motivated by the rigorous probing of their electronic structure and dynamics through the Regge pole-calculated R-T minima and SRs, revealing their sensitivity [1]. Also importantly, the flipping over of the deep R-T minimum in the Bk TCSs to a SR very close to threshold occurs in the metastable TCS of the Cf atom; see the orange curves in Fig. 4. The results of probing their electronic structure and dynamics by the experiment [31] as well as our validation of the observation [1] should help in the measurements of the EAs of Bk and Cf. In Table 1 we have presented our Regge pole anionic BEs for both Bk and Cf and compared them with the existing theoretical EAs; these are riddled with uncertainty and lack reliability. The same conclusion applies to the data of the Am atom.

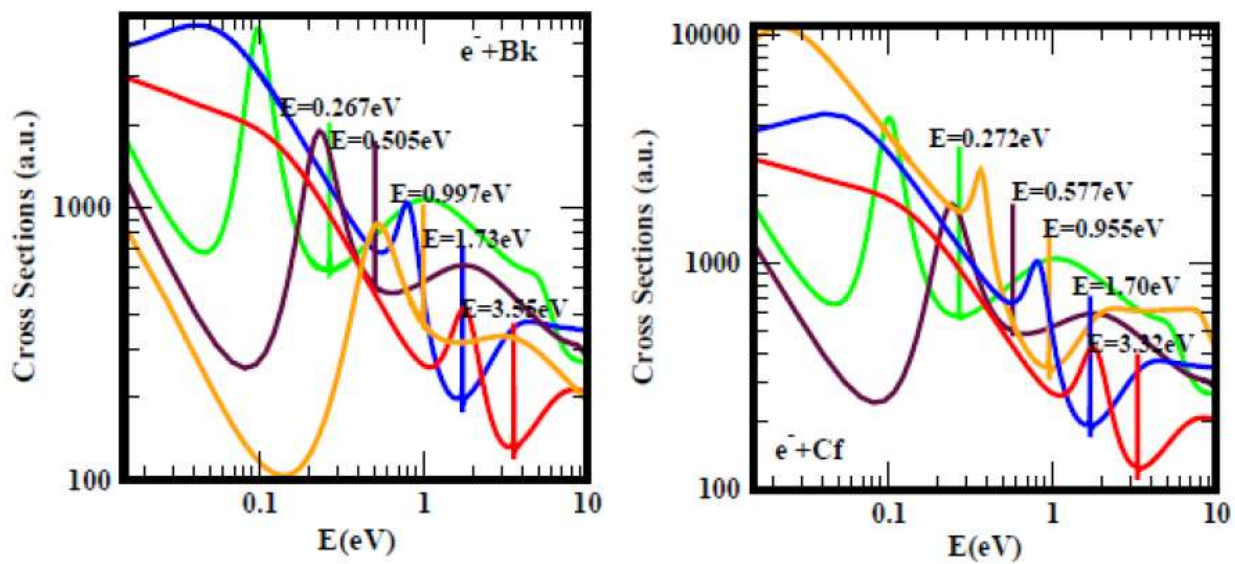

**Figure 4:** Total cross sections (a.u.) for electron elastic scattering from atomic Bk (left panel) and Cf (right panel) are contrasted. For both Bk and Cf the red, blue and orange curves represent TCSs for the ground and the two metastable states, respectively while the brown and the green curves correspond to excited states TCSs. The dramatically sharp resonances in the TCSs of both figures correspond to the Bk⁻ and Cf⁻ negative-ions formation during the collisions. Note that the flip over of the near threshold R-T minimum from the Bk TCSs to a SR very close to threshold in the Cf TCSs occurs here.

### 3.5 Fm and Lr atoms

There are no measured EAs for these atoms; therefore reliable predictions of their EAs are essential. As the size of the actinide atoms approaches that of the Lr atom, their electronic structure becomes less complicated and the theoretical EAs are expected to become less uncertain. Figure 5 compares the TCSs of the Fm and Lr atoms, selected because of the availability of several theoretical EAs to compare with the Regge pole-calculated BEs. The TCSs of both atoms are characterized by well delineated dramatically sharp resonances, representing in each atom a ground, two metastable and two excited states BEs, see also Table 1. Worth remarking here is that in both the TCSs of Fm and Lr the sharp resonances of interest here, although well-delineated from each other, appear close to SRs; this can create problems in the identification of the BEs as seen in Fig. 5.

For Fm we see that the Regge pole BEs of the highest excited anionic state (0.268eV) and the second excited anionic state (0.623eV) agree rather well with the existing theoretical EA values of 0.354eV [24] and 0.597eV [24], respectively. Importantly, the reason why these theoretical EAs differ from each other is that they correspond to different anionic states; this demonstrates the need for rigorous values of the EAs for Fm and the other actinide atoms in general to guide measurements and/or calculations. We note here that the Fm TCSs still exhibit fullerene behavior, while in Lr the fullerene behavior has completely disappeared (see green curves in both figures of Fig. 5). This almost atomic behavior of the Lr TCSs makes the Lr much easier to handle theoretically as seen below. Also, the deep R-T minima now appear as SRs near threshold in both the Fm and Lr TCSs (orange and brown curves, respectively), having flipped over for the first time in the TCSs of Bk to those of Cf.

The importance of the Lr TCSs is two-fold: 1) The electronic structure of Lr is relatively simple and various sophisticated theoretical methods have calculated the EA of Lr, see Table 1. 2) On comparing the TCSs of the Fm and Lr atoms, we see that the characteristic SRs appear very near threshold in both TCSs. Since the Lr atom is the last of the actinide series, sophisticated theoretical methods can be expected to obtain better values for the EA of Lr. Indeed, as in atomic Fm the Lr TCSs are seen to be characterized by well delineated ground (3.88eV), two metastable (1.92eV; 1.10eV) and two excited (0.321eV; 0.649eV) states anionic BEs. Once the determination has been made regarding what anionic state BE was measured/calculated, then by comparison with the above values, an unambiguous EA determination can be obtained as was done with the cases of Au, Pt, At and the fullerene molecules in Table 1. However, several calculated EAs are available; we will attempt to make sense of their meaning. Firstly, the 0.310eV [25], 0.295eV [23] and the Abs (-0.313eV) [24] EAs can safely be identified with the Regge pole highest excited state BE (0.321eV) and secondly, the EA values of 0.465eV [23] and 0.476eV [26] could probably be identified with the Regge pole BE of the second excited state 0.649eV; they could also correspond to the nearby shape resonance at 0.451eV, see Fig. 5.

Indeed, from Table 1 it is clear that for the actinide atoms the various sophisticated theoretical methods calculate only the BEs of the formed negative-ions in excited states and equate them with the EAs. There is nothing wrong with this viewpoint, except that a rigorous definition of the EA is required for consistency throughout the tabulated atoms and fullerenes in Table 1. This would then mean using the BEs in the column BEs, EXT-1 for the corresponding EAs of all the atoms and fullerene molecules in Table 1. Unfortunately, this would contradict the established meaning of the EAs as found for the Au, Pt, and At atoms as well as for the $C_{60}$ and $C_{70}$ fullerenes. It is further noted here that for the carefully measured and calculated EA of At, various sophisticated theoretical EAs [32-35] agree excellently among themselves and with the measured EA as well as with the Regge pole ground state anionic BE and not with the metastable or the excited state anionic BEs. Unfortunately, it is a formidable task for most theoretical methods to calculate the BEs of the metastable states, let alone the ground states BEs of the tabulated systems in Table 1.

These results are also important in guiding sophisticated theoretical methods on the importance of the polarization interaction. This has been demonstrated unequivocally in the TCSs calculation of the Bk and Cf atoms, particularly in the At atom. When the "b" parameter of the Eq. (2) was 0.042 we obtained the ground state anionic BE value of 2.51eV [56], which was close to the then known theoretical EA value of 2.80eV [57]. However, a careful refinement of "b" to 0.04195 yielded the ground state BE value of 2.42eV, in excellent agreement with the measured EA of 2.416eV [7] and the theoretical values [32 - 35]. Here it is noted that the small Im ($\lambda$) decreased from $1.07 \times 10^{-5}$ to $8.9 \times 10^{-6}$ indicative that the BE value of 2.42eV corresponds to the longest-lived resonance as expected (see the importance of the use of Im ($\lambda$) in the Regge pole analysis in Ref. [40]). Indeed, a scientific will is needed to respond critically to the question: why are the EAs of Au, Pt and At as well as those of the $C_{60}$ and $C_{70}$ fullerene molecules identified with the BEs of the formed negative ions in the ground states while for Th and U as well as for the other actinide atoms the EAs correspond to the Regge pole-calculated BEs of excited states? Notably, the rigorous Regge pole BEs are available to guide the EA measurements, regardless of their interpretation, namely whether they are viewed as corresponding to the BEs of electron attachment in the ground or excited states.

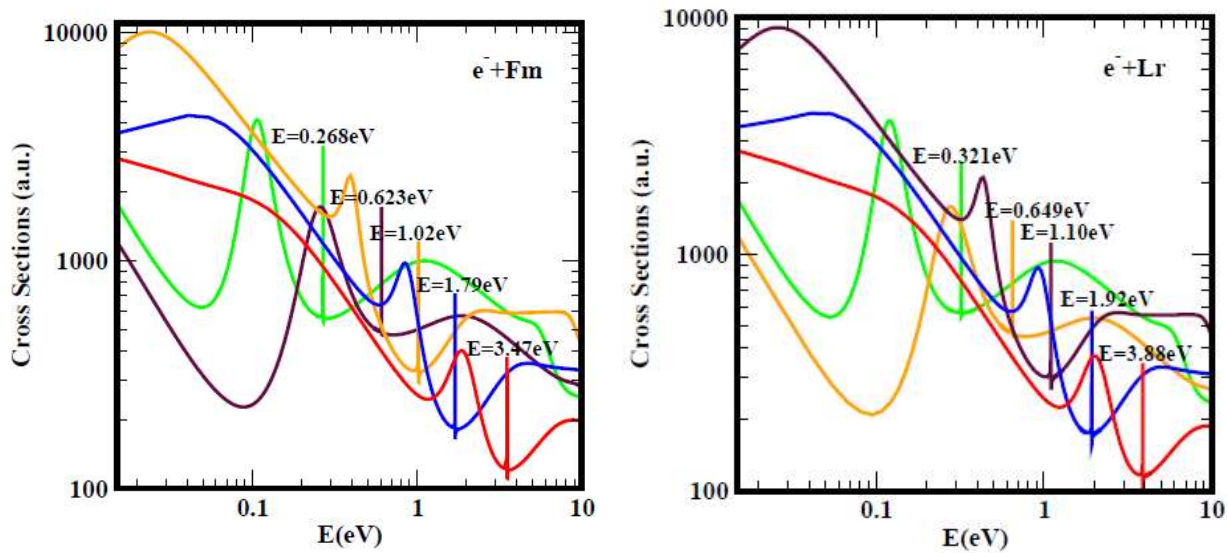

**Figure 5:** Total cross sections (a.u.) for the actinide atoms Fm (left panel) and Lr (right panel). In both panels the red curves represent the ground states. In Fm the blue and the orange curves and in Lr the blue and the brown curves are the metastable TCSs while the brown and the green curves in Fm and the orange and the green curves in Lr represent the excited states TCSs. The energy positions of the dramatically sharp lines correspond to electron BEs of the formed negative ions during collisions. The orange curve in the TCSs for Fm and the brown curve in those for Lr correspond to the polarization-induced TCSs with the deep R-T minima flipped over to SRs close to threshold (see also Fig. 4 above).

### 3.6 Relativistic effects in Electron Affinity calculations

The EA provides a stringent test of theoretical calculations when the calculated EAs are compared with those from reliable measurements. Unfortunately, low-energy electron interactions with heavy multi-electron atoms and fullerene molecules are characterized generally by the presence of many intricate and diverse electron configurations. These lead to computational complexity that, for a long time, made it virtually impossible for sophisticated theoretical methods to reliably predict the electron binding energies of the formed negative ions during collisions. Thus the electron affinities calculated using many structure-based theoretical methods tend to be riddled with uncertainties making them difficult to interpret, particularly for the actinide atoms, see Table 1.

One of the most important and revealing investigations of the importance of Regge Trajectories in low-energy electron collisions using the ABF, Eq. (2), potential was carried out by Thylwe [41]. For the Xe atom Regge Trajectories calculated using the Dirac Relativistic and non-Relativistic methods were contrasted near threshold and found to yield essentially the same Re $\lambda(E)$ when the Im $\lambda(E)$ was still very small, see Fig. 2 of [41]. This implies the insignificant difference between the Relativistic and non-Relativistic calculations at low-energy electron scattering, the condition of our calculations here.

As indicated in the introduction, most of the sophisticated theoretical calculations of the EAs include relativistic effects at various levels of approximations, see for example their comparisons in the calculation of the EA of At in Ref. [34]. Since many of these methods are tailored to reproduce the measurements very well, it is difficult to determine what essential physics is incorporated in the calculation of the EAs. For instance, Wesendrup et al [58] carried out large-scale fully Relativistic Dirac-Hartree Fock and MP2 as well as nonrelativistic pseudo potential calculations and obtained the EAs of Au as 2.19eV and 1.17eV, respectively. Also, Cole and Perdew [27] employed Relativistic and nonrelativistic calculations obtaining the EA of atomic Au as 2.5eV and 1.5eV, respectively.

**Table 1:** Negative-ion binding energies (BEs) and ground states Ramsauer-Townsend (R-T) minima, all in eV extracted from TCSs of the atoms and the fullerene molecules $C_{60}$ and $C_{70}$. They are compared with the measured electron affinities (EAs) in eV. GRS, MS-$n$ and EXT-$n$ ($n$=1, 2) refer respectively to ground, metastable and excited states. Experimental EAs, EXPT and theoretical EAs, Theory are also included. The numbers in the square brackets are the references.

| System Z | BEs GRS | BEs MS-1 | BEs MS-2 | EAs EXPT | BEs EXT-1 | BEs EXT-2 | R-T GRS | BEs/EAs Theory | EAs RCI[23] | EAs GW[24] |
|---|---|---|---|---|---|---|---|---|---|---|
| Au 79 | 2.26 | 0.832 | - | 2.309[2] 2.301[3] 2.306[4] | 0.326 | - | 2.24 | 2.50[27] 2.19[58] 2.313[59] 2.263[60] | - | - |
| Pt 78 | 2.16 | 1.197 | - | 2.128[2] 2.125[5] 2.123[6] | 0.136 | - | 2.15 | 2.163[60] | - | - |
| At 85 | 2.42 | 0.918 | 0.412 | 2.416[7] | 0.115 | 0.292 | 2.43 | 2.38[32] 2.42[33] 2.412[34] 2.45 [35] | - | - |
| $C_{60}$ | 2.66 | 1.86 | 1.23 | 2.684[8] 2.666[9] 2.689[10] | 0.203 | 0.378 | 2.67 | 2.57[61] 2.63[62] 2.663[63] | - | - |
| $C_{70}$ | 2.70 | 1.77 | 1.27 | 2.676[9] 2.72[11] 2.74[12] | 0.230 | 0.384 | 2.72 | 3.35[64] 2.83[64] | - | - |
| Nb 41 | 2.48 | 0.902 | - | 0.917[13] 0.894[14] | 0.356 | - | 2.47 | 0.82 [27] 0.99 [28] | - | - |
| Eu 63 | 2.63 | 1.08 | - | 0.116[16] 1.053[17] | 0.116 | - | 2.62 | 0.117[22] 0.116[40] | - | - |
| Tm 69 | 3.36 | 1.02 | - | 1.029[18] | 0.016 | 0.274 | 3.35 | - | - | - |
| Hf 72 | 1.68 | 0.525 | - | 0.178[15] | 0.017 | 0.113 | 1.67 | 0.114[52] 0.113[53] | - | - |
| Th 90 | 3.09 | 1.36 | 0.905 | 0.608 [19] | 0.149 | 0.549 | 3.08 | 0.599 [19] | 0.368 | 1.17 |
| U 92 | 3.03 | 1.44 | - | 0.315[20] 0.309[21] | 0.220 | 0.507 | 3.04 | 0.175 [55] 0.232[21] | 0.373 | 0.53 |
| Am 95 | 3.25 | 1.58 | 0.968 | N/A | 0.243 | 0.619 | 3.27 | - | 0.076 | 0.103 0.142 |
| Bk 97 | 3.55 | 1.73 | 0.997 | N/A | 0.267 | 0.505 | 3.56 | - | 0.031 | -0.276 -0.503 |
| Cf 98 | 3.32 | 1.70 | 0.955 | N/A | 0.272 | 0.577 | 3.34 | - | 0.010 0.018 | -0.777 -1.013 |
| Fm 100 | 3.47 | 1.79 | 1.02 | N/A | 0.268 | 0.623 | 3.49 | - | - | 0.354 0.597 |
| Lr 103 | 3.88 | 1.92 | 1.10 | N/A | 0.321 | 0.649 | 3.90 | 0.160[25] 0.310[25] 0.476[26] | 0.295 0.465 | -0.212 -0.313 |

These EAs should be compared with the nonrelativistic Regge pole-calculated BE value of 2.263eV [1, 60] and the measured EAs presented in Table 1. Accordingly, it can be safely concluded that in the energy regime, $0.0 \leq E \leq 10.0$eV the essential physics embedded in the Regge pole method such as electron-electron correlations and core-polarization interaction is adequate for the reliable prediction of the EAs of multi-electron atoms and fullerene molecules. Importantly, an impressive agreement with the measured EAs of Au [2, 3, 4] has been obtained by the Relativistic Coupled Cluster Calculations with Variational Quantum Electrodynamics [59].

The calculation of the EA of Nb ($Z = 41$) in [27] used a gradient-corrected exchange correlation functional and a Nb scalar-relativistic core, obtaining the EA value of 0.82eV, which compares very well with the Regge pole metastable BE value of 0.902eV and the measured EA values of 0.917eV [13] and 0.894eV [14]. These results for Nb further demonstrate the great need to ascertain precisely the state whence the photodetachment process originates. The Eu atom, with a relatively high Z of 63, but a small measured EA of 0.116eV [16], provides a stringent test of the nonrelativistic Regge pole method when its BE value of 0.116eV [40] is contrasted with the MCDF-RCI calculated EA value of 0.117eV [22]. The theoretical results were calculated around 2009 while the experimental EA was measured in 2015. For the highly radioactive At atom the recently measured EA [7], which employed the Coupled-Cluster method, agreed excellently with the Regge pole BE and the EAs from various sophisticated theoretical calculations, including the Multiconfiguration Dirac Hartree-Fock values [32-35], see Table 1 for comparisons. Furthermore, in [32, 34] extensive comparisons among various sophisticated theoretical EAs have been carried out as well.

## 4. Summary and Conclusion

For all the multi-electron atoms and the fullerene molecules considered in this paper, we extracted from the Regge pole-calculated TCSs rigorous and unambiguous ground, metastable and excited states negative-ion BEs of the formed anions during the collision. We then compared our BEs with the existing measured and/or calculated EAs as shown in Table 1. We found that for the Au, Pt, and At atoms as well as the $C_{60}$ and $C_{70}$ fullerene molecules our ground state anionic BEs matched excellently with the measured EAs, implying that the measured EAs of these systems correspond to the BEs of the electron when it is attached in the ground states of the formed negative ions during the collision. However, for the lanthanide atom Eu, our excited state BE is in outstanding agreement with both the recently measured and the MCDF-RCI calculated EA values. For both the Eu and Tm atoms very good agreement between the Regge pole metastable BEs and the previously measured EAs [17, 18] has been realized. Overall, our excited states BEs are closer to the measured and/or calculated EAs for the Hf and the actinide atoms. This implies that for these atoms the measured and/or calculated EAs correspond to the BEs of the electron when it is attached in the excited states of the formed negative ions, contrary to the cases of the Au, Pt and At atoms as well as the $C_{60}$ and $C_{70}$ fullerene molecules.

Indeed, the meaning of the EAs in these multi-electron systems is both ambiguous and confusing; it lacks uniformity as well. Unfortunately, a number of the existing sophisticated theoretical methods used to guide the measurements tend to produce EAs that are riddled with uncertainty and currently generate BEs for only the excited states, see Table 1. We recommend that, for progress of the field, the various sophisticated theoretical methods use our robust and rigorous BEs to generate the electronic wave functions and the fine structure energies for the various atoms. As an example, the MCDF-RCI calculations determined that the $5d^26s^26p$ $J = 5/2$ is the only bound state of Hf$^-$ and the EA of Hf is 0.114eV [52]. This value agrees excellently with the Regge pole-calculated BE value of 0.113eV but it is for an excited state. Clearly, these results demonstrate the utility of the Regge pole anionic BEs in guiding the building of the CI wave functions.

In conclusion, the existing measured/calculated EAs of the multi-electron atoms and fullerene molecules considered in this paper have been identified with the Regge pole-calculated electron BEs when it is attached in the ground, metastable and excited states of the formed negative-ions during the collisions. The general lack of unambiguous and uniform meaning of the EAs of the multi-electron atoms and fullerene molecules is of great concern to us. Promising, however is the recent accomplishment in the measurement of the EA of the highly radioactive At atom, supported by various sophisticated theoretical calculations, including the Regge pole-calculated ground state anionic BE. It is hoped that the actinide atoms will be subjected to similar investigations as in [32, 34] for unambiguous and reliable EAs. The great strength of the Regge pole analysis is in its use of Im λ(E) to differentiate among the ground, metastable and excited negative ion states, with the the ground state having the smallest Im λ(E), indicative of the longest-lived negative-ionic state.

## Author Contributions

Z.F. and A.Z.M. are responsible for the conceptualization, methodology, investigation, formal analysis and writing of the original draft, as well as rewriting and editing. A.Z.M. is also responsible for securing the funding for the research. All authors have read and agreed to the published version of the manuscript.

## Acknowledgments

Research was supported by the U.S. DOE, Division of Chemical Sciences, Geosciences and Biosciences, Office of Basic Energy Sciences, Office of Energy Research, Grant: DE-FG02-97ER14743. The computing facilities of National Energy Research Scientific Computing Center, also funded by U.S. DOE are greatly appreciated.

## Conflicts of Interest

The authors declare no conflict of interest or state.